\documentclass[fleqn,usenatbib]{mnras}

\usepackage{newtxtext,newtxmath}
\usepackage{chemformula}
\usepackage{longtable}
\usepackage{tabularray}
\usepackage{tabularx}

\usepackage[T1]{fontenc}
\usepackage{ulem}

\DeclareRobustCommand{\VAN}[3]{#2}
\let\VANthebibliography\thebibliography
\def\thebibliography{\DeclareRobustCommand{\VAN}[3]{##3}\VANthebibliography}

\usepackage{graphicx}	
\usepackage{amsmath}	

\title[B-fields in cometary clouds]{The magnetic fields around the cometary globules, L328, L323 and L331}

\author[Kumar et al.]{
Siddharth Kumar$^{1}$\thanks{E-mail: ksiddharth@iisc.ac.in},
Archana Soam$^{2}$, Nirupam Roy$^{1}$
\\
$^{1}$Department of Physics, Indian Institute of Science, Bangalore 560012, India\\
$^{2}$Indian Institute of Astrophysics, II Block, Koramangala, Bengaluru 560034, India\\
$^{3}$Department of Physics, Indian Institute of Science, Bangalore 560012, India\\
}

\date{Accepted 2023 June 09. Received 2023 June 08; in original form 2022 December 19}

\pubyear{2023}

\begin{document}
\label{firstpage}
\pagerange{\pageref{firstpage}--\pageref{lastpage}}
\maketitle

\begin{abstract}
This work presents the magnetic field geometry in a complex of three cometary (with head-tail morphology) globules, namely LDN 323, LDN 328, and LDN 331, using R-band polarization measurements of background stars. These observations were combined with a $Planck$ sky survey to study the large-scale morphology of the magnetic fields in the region. The distances of the target stars were adopted from the Gaia catalog. The variation of degree of polarization and polarization position angle with distances of stars is analyzed. The field geometry is mostly found to follow the cometary shape of the cloud, with some randomness at certain locations. For studying the correlation between cloud morphology and magnetic field orientations, a modified version of the Histogram of Relative Orientation analysis was employed. 

\end{abstract}




\section{Introduction}

The birth of a massive star ionises the surrounding region. This process has been observed in a number of rapidly expanding HII regions. The shock generated by this ionization heating will converge into the cloud, compressing it into single or multiple cores, which may eventually collapse to form stars. Several patches of obscuration can be found on the outskirts of relatively mature HII regions.\citep{1983ApL....23..119Z, 1991ApJS...77...59S, 1994ApJS...92..163S, 2010A&A...523A...6D, 2010A&A...518L..81Z, 2012AJ....144..173D, 2012A&A...542L..18S}.The central ionising source(s) generated by recombination radiation from the ionization front (I-front) can cause some of these patches to have a bright rim on the surface facing the ionising source; these patches are then referred to as "bright-rimmed clouds" (BRCs). A number of these BRCs have been discovered to be associated with Infrared Astronomical Satellite (IRAS) point sources. 44 in the northern and 45 in the southern hemisphere have been catalogued in \citet{1991ApJS...77...59S} and \citet{1994ApJS...92..163S}.These BRCs are classified into three types based on their length-to-width ratios, with type C being the most elongated \citep{1991ApJS...77...59S, 1994ApJS...92..163S}. A large number of these BRCs show signs of ongoing star formation, which could be triggered by the formation of an ionization front \citep{1989ApJ...342L..87S, 1995ApJ...455L..39S, 2002AJ....123.2597O, 2011MNRAS.415.1202C, 2012PASJ...64...96H, 2012Ap.....55..471M}. For these reasons, BRCs are an ideal testing environment for studying the radiation-driven implosion process, which has been proposed as one of the mechanisms responsible for cloud compression and subsequent star formation \citep{1983A&A...117..183R, 1989ApJ...346..735B, 1994A&A...289..559L}.


\vspace{10pt}
We have made significant progress in the development of several semi-analytical and numerical modeling studies to understand the details of the radiation-driven implosion (RDI) of photoionized clumps and their subsequent acceleration away from the ionising source over the last few decades. \citep[e.g., ][]{2003MNRAS.338..545K, 2009ApJ...694L..26G, 2009ApJ...692..382M, 2011ApJ...736..142B, 2012MNRAS.420..562H, 2013MNRAS.431.3470H, 2015MNRAS.450.1017K}.The first 2-D radiation-magnetohydrodynamics (R-MHD) calculation to investigate the influence of magnetic fields on the evolution of the expanding HII region was done by \citet{2007Ap&SS.307..179W}. Further investigation revealed significant differences in the values of the allowed velocities of the propagation of the ionization fronts between magnetized and non-magnetized cases by solving the continuity equation for the simple cases of a magnetic field parallel \citep{1998A&A...331.1099R} and oblique \citep{2000MNRAS.314..315W} to the ionization front. Along with the morphology of the magnetic fields, the strength of these magnetic fields must be studied.  \citet{2009MNRAS.398..157H}  ran 3-D R-MHD simulations with varying magnetic field strengths and orientations. The results of these simulations revealed that the magnetic field orientation played an important role in shaping the morphology of the clouds: a strong perpendicular (to the direction of ionising radiation) magnetic field orientation results in a flat, plate-like structure, whereas an oblique initial field results in a comma-shaped cloud morphology. Mapping magnetic fields in these areas are critical for these investigations.\\

\begin{figure}
\includegraphics[width=\columnwidth]{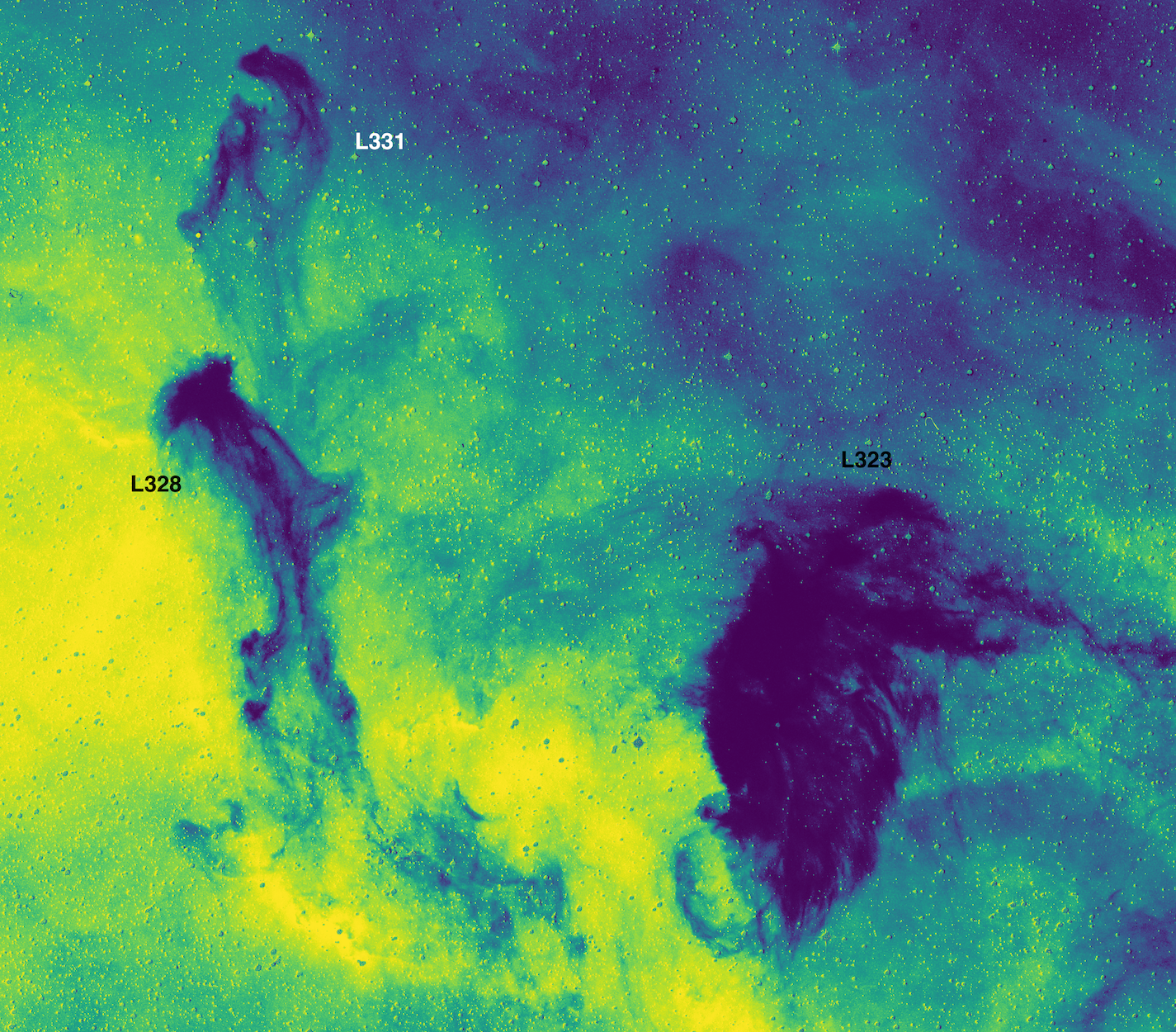}
\caption{The figure above shows the H${\alpha}$ image of the region, showing all three clouds L323, L328
and L331.}
\label{fig:Halpha}
\end{figure}

One of the most important studies in a polarimetric investigation is the mapping of the interstellar magnetic field. The observations of starlight polarization by absorption in visible and near-infrared wavelengths are an essential tool in tracing the plane of the sky component of the magnetic field. \citep{1949Sci...109..166H, 1949ApJ...109..471H, Hiltner1949}. Polarization of starlight in the visible and near-infrared is caused by the dichroic absorption of starlight by non-spherical dust grains. These spinning dust grains are aligned such that their minor axis is parallel to the rotation axis and the local magnetic field. Such alignment leads to the dichroism of extinction along the long axis, resulting in a net polarization parallel to the magnetic field. This is one of the most common and effective methods of deriving the magnetic field morphology towards molecular clouds in different environments \citep[e.g., ][]{1976AJ.....81..958V, 1987ApJ...319..842H, 1992MNRAS.257...57B, 2002ApJS..141..469P, 2011ApJ...741...21C, 2011ApJ...734...63S, 2013MNRAS.432.1502S, 2014A&A...565A..94B, 2015ApJ...798...60K, 2015ApJ...803L..20S}. The same aligned dust grains can also be observed in far-infrared and sub-millimeter wavelengths emitting polarized thermal radiation \citep[e.g.,][]{1982MNRAS.200.1169C, 1998ApJ...502L..75R, 2000ApJS..128..335D, 2010ApJS..186..406D, 2012ApJS..201...13V, 2014ApJS..213...13H}. As the polarization of thermal emission is largest along its longer axis, this results in a net polarization of radiation emitted by dust grains perpendicular to the local magnetic field. These polarization angles then must be rotated by 90$^o$ to get the magnetic field orientation. Even after many advances in the study of interstellar dust grain alignment, the exact process still remains a mystery. \citep{2007JQSRT.106..225L, 2015ARA&A..53..501A}. However, the radiative torque mechanism, first proposed by \citet{Dolginov}, has been widely considered to be the most successful one in explaining such alignments in various environments, \citep[e.g., ][]{2014MNRAS.438..680H, 2015MNRAS.448.1178H, 2015ARA&A..53..501A}. For this study we have chosen LDN 328 (hereafter L328), LDN 323 (hereafter L323), and LDN 331 (hereafter L331) seen in figure \ref{fig:Halpha}, three dark nebulae in close proximity to each other for our study. Close proximity of the three clouds and similar orientation of the head-tail morphology suggest a common external ionizing source. There have been many studies done previously on L328 but not on the whole region including L323 and L331 together in the context of dust polarization. This paper has been divided into four sections. The first section provides a general introduction to the paper; the second section describes the data collection process; the third section presents all of the results, and the fourth section provides the paper's conclusions.  \\

\section{Observations and data reduction}
The main data used in this paper was observed at Sampurnanand telescope in ARIES, India with the active optical polarimeter, AIMPOL. It was observed in optical (R-band) on May 15 and May 16 2013 using an achromatic half-wave plate (HWP) modulator and a Wollaston prism beam splitter. A Johnson R$_{KC}$ filter was used to separate out the R-band from optical observation. The observation was carried out on a 1024 $\times$ 1024 pixel$^2$ CCD chip (Tektronic TK1024) and when combined with a camera of 8$^{\prime}$ field of view (FOV) falls on the central 325 $\times$ 325 pixels$^2$. The CCD plate had a plate scale of 1.48$^{\prime\prime}/$pixel and the field of view is $\sim 8^{\prime}$ in diameter. The CCD had the read-out noise and gain of 7.0 $e^{-}$  and 11.98 $e^{-}$/ADU respectively. The full width half maximum (FWHM) varies from 2 to 3 pixels. The benefit of using the dual-beam polarizing prism is that it allows us to measure the polarization by simultaneously imaging both orthogonal polarization states onto the detector. The half-wave plate fast axis and the axis of the Wollaston prism were kept normal to the optical axis of the system. We can write the intensities of extraordinary and ordinary beams as  

\begin{equation}
I_{e}(\alpha)=\frac{I_{unpol}}{2}+I_{pol}\times cos^{2}(\theta-2\alpha)
\end{equation}

\begin{equation}
I_{o}(\alpha)=\frac{I_{unpol}}{2}+I_{pol}\times sin^{2}(\theta-2\alpha)
\end{equation}

where, {\it $I_{o}$}, {\it $I_{e}$} are the fluxes of ordinary and extraordinary beams and $I_{pol}$ and $I_{unpol}$ are the polarised and unpolarised intensities respectively.  $\theta$ and $\alpha$ are the position angles for the polarization vector and the half-wave plate fast axis respectively, measured with respect to the Wollaston prism. The reduction of the data was done using the standard aperture photometry in the IRAF package and the fluxes of ordinary ({\it $I_{o}$}) and extraordinary ({\it $I_{e}$}) beams for the observed sources were extracted with a good signal-to-noise ratio. The ratio {\it {R($\alpha$)}} is given by

 \begin{equation}
 R(\alpha) = \frac{\frac{{I_{e}}(\alpha)}{{I_{o}}(\alpha)}-1} {\frac{I_{e}(\alpha)} {I_{o}(\alpha)}+1} =  P\times cos(2\theta - 4\alpha)
\end{equation}

where {\it P} and $\theta$ are the fraction of total linearly polarised light and the polarization angle of the plane of polarization. {\it $\alpha$} here is the position of the fast axis of the HWP at  $0^o$, $22.5^o$, $45^o$, and $67.5^o$. These angles correspond to four normalised Stokes parameters namely q[R($0^o$)], u[R($22.5^o$)], $q_{1}$[R($45^o$)] and $u_{1}$[R($67.5^o$)]. The errors in the normalised Stokes parameters ($\sigma_R$)($\alpha$)($\sigma_q$, $\sigma_u$, $\sigma_{q1}$, $\sigma_{u1}$) can be estimated using the relation \citep{1998A&AS..128..369R}

\begin{equation}
\sigma_R(\alpha)= \frac{\sqrt{N_{e}+N_{o}+2N_{b}}}{N_{e}+N_{o}}
\end{equation}

\begin{figure}
\includegraphics[width=\columnwidth]{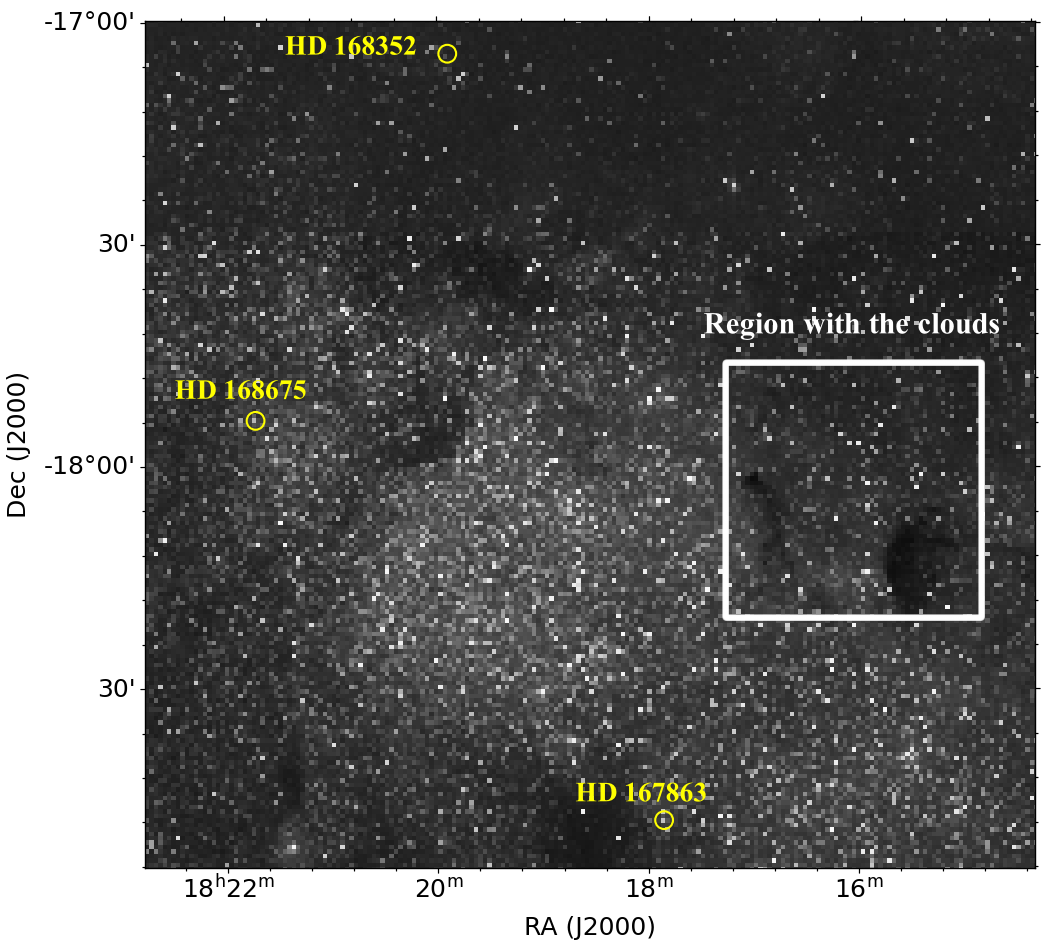}
\caption{The figure above shows the closest B type stars within 2$^o$ of the cloud structures shown on a DSS2 red band image of the region}
\label{fig:star_plots}
\end{figure}

\begin{figure*}
\includegraphics[width=0.85\textwidth]{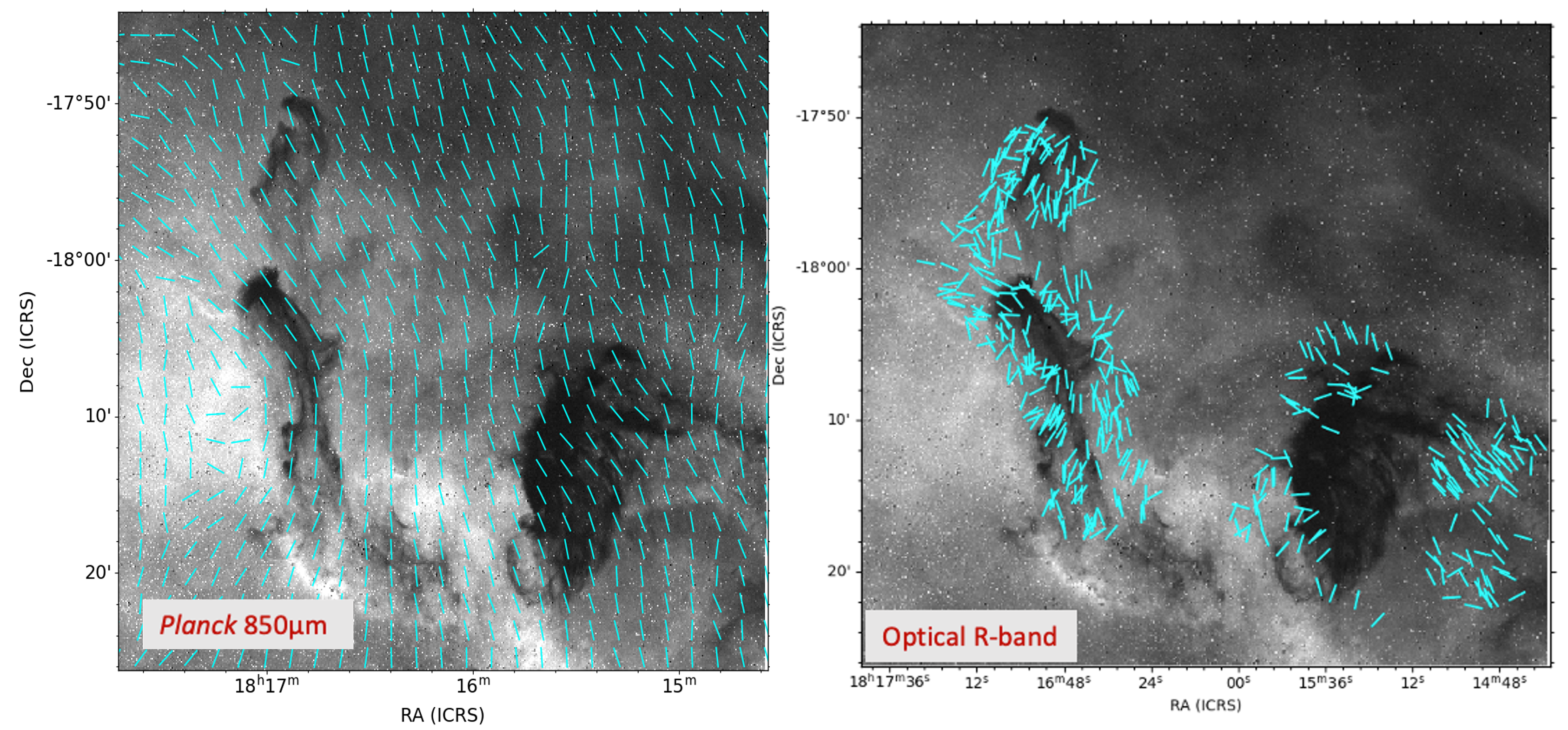}
\caption{The figure above shows the plane of the sky magnetic field inferred from $Planck$ polarization vectors on the left and optical polarization vectors on the right, both plotted on an R-band subtracted H$\alpha$ image.}
\label{fig:plank_optical}
\end{figure*}


where $N_{o}$ and $N_{e}$ are the counts of ordinary and extraordinary beams, respectively, and $N_{b}$[= ({$N_{be}$}$+${$N_{bo}$})/2] is the average background counts around the extraordinary and ordinary rays.


\section{Results and discussion}


\subsection{Cloud morphology}
\begin{figure*}
\includegraphics[width=0.6\textwidth]{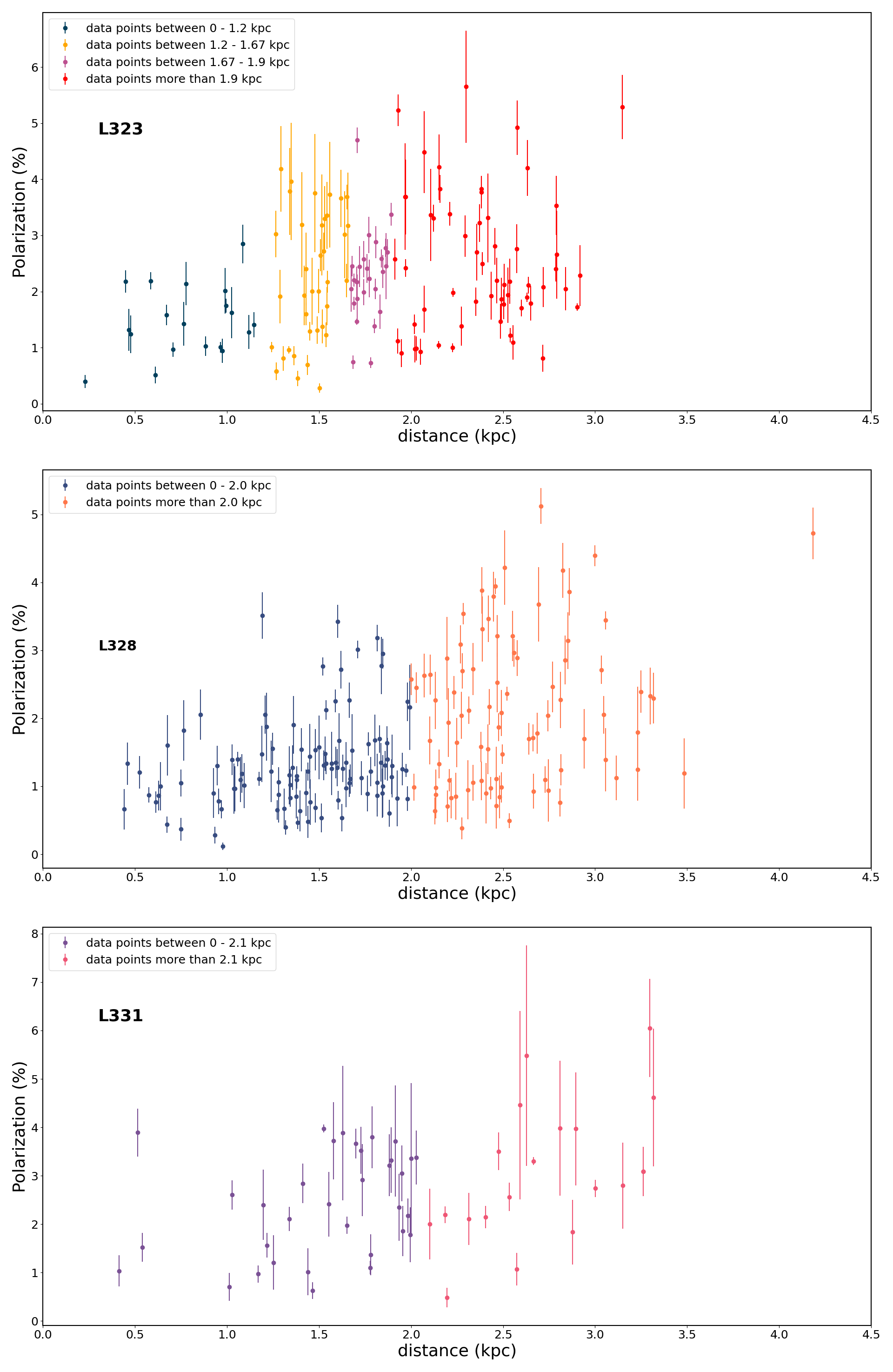}
\caption{The figure above shows the polarization percentage as a function of distance obtained from Bailer-Jones 2021 for the clouds (from top to bottom) L323, L328, and L331 }
\label{fig:p_vs_d}
\end{figure*}

\begin{figure*}
\includegraphics[width=0.6\textwidth]{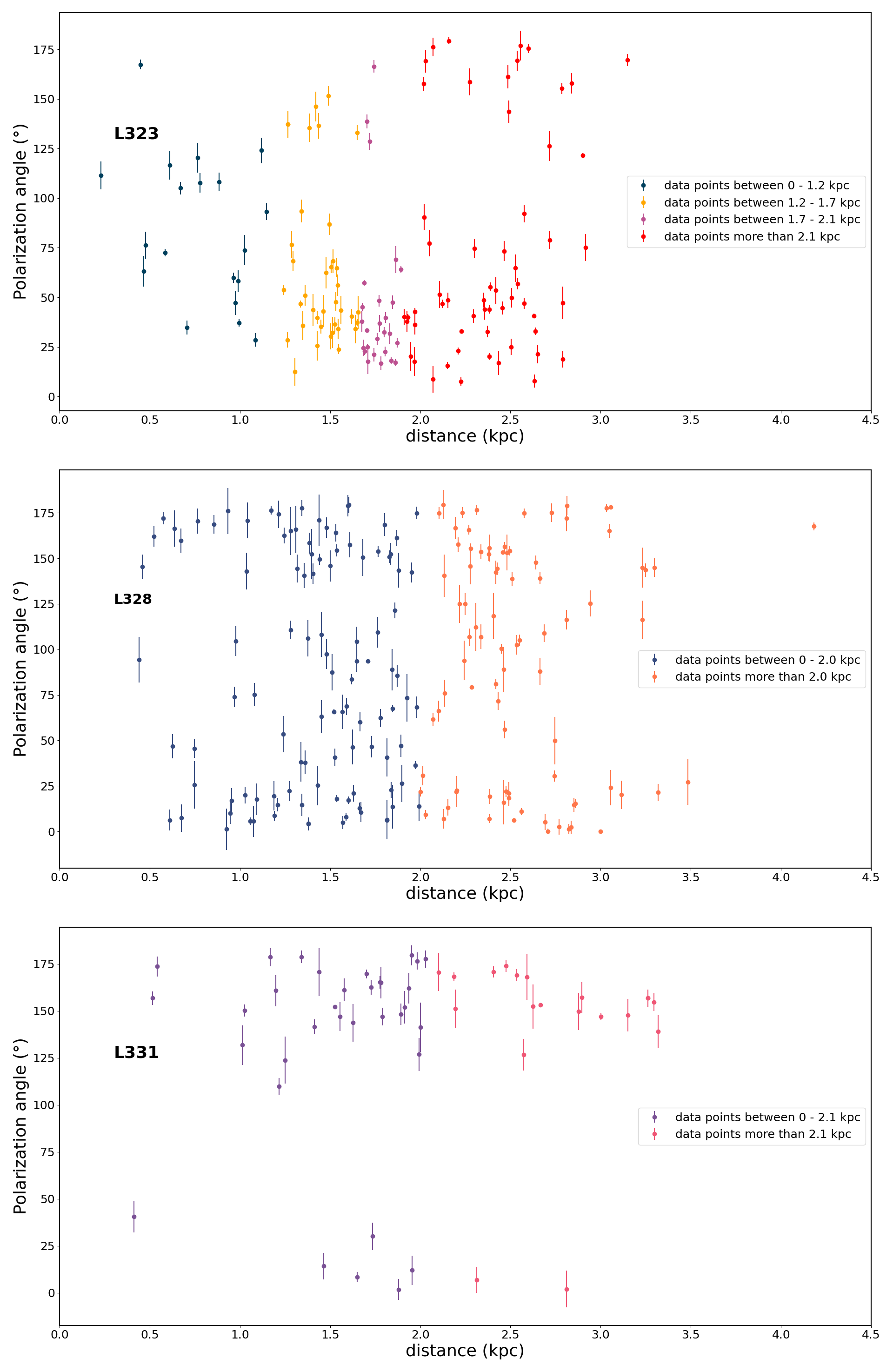}
\caption{The figure above shows polarization angle as a function of distance obtained from Bailer-Jones 2021 for the clouds (from top to bottom) L323, L328 and L331}
\label{fig:pa_vs_d}
\end{figure*}

\begin{figure*}
\includegraphics[width=1\textwidth]{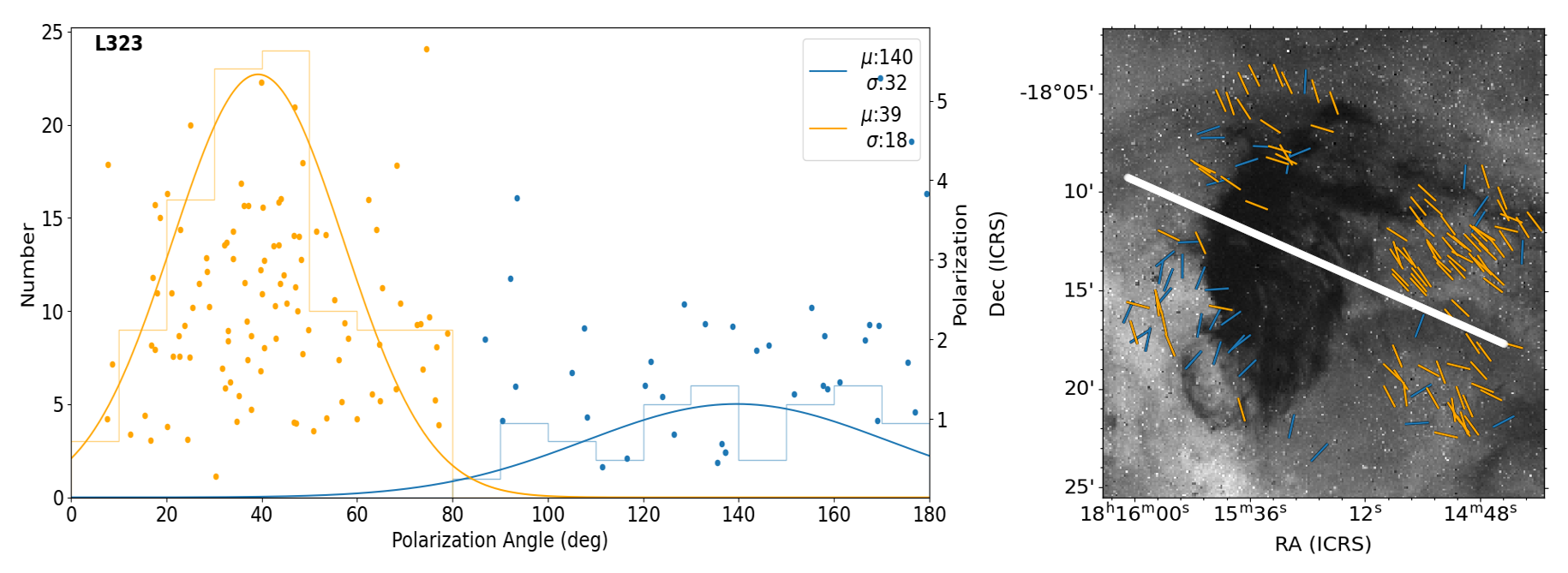}
\caption{The figure above shows the distribution of polarization angles for L323 with a Gaussian fit to the polarization angle distribution on the left. The distribution has been divided into smaller distributions \textbf{I} and \textbf{II} colored as yellow and blue respectively. The colored dots represent the polarization of stars, where larger dots show higher polarization and vice-versa. On the right we have distributions \textbf{I} and \textbf{II} plotted on the cloud with the major axis of the cloud plotted in white.}
\label{fig:hist_L323}
\end{figure*}

\begin{figure*}
\includegraphics[width=1\textwidth]{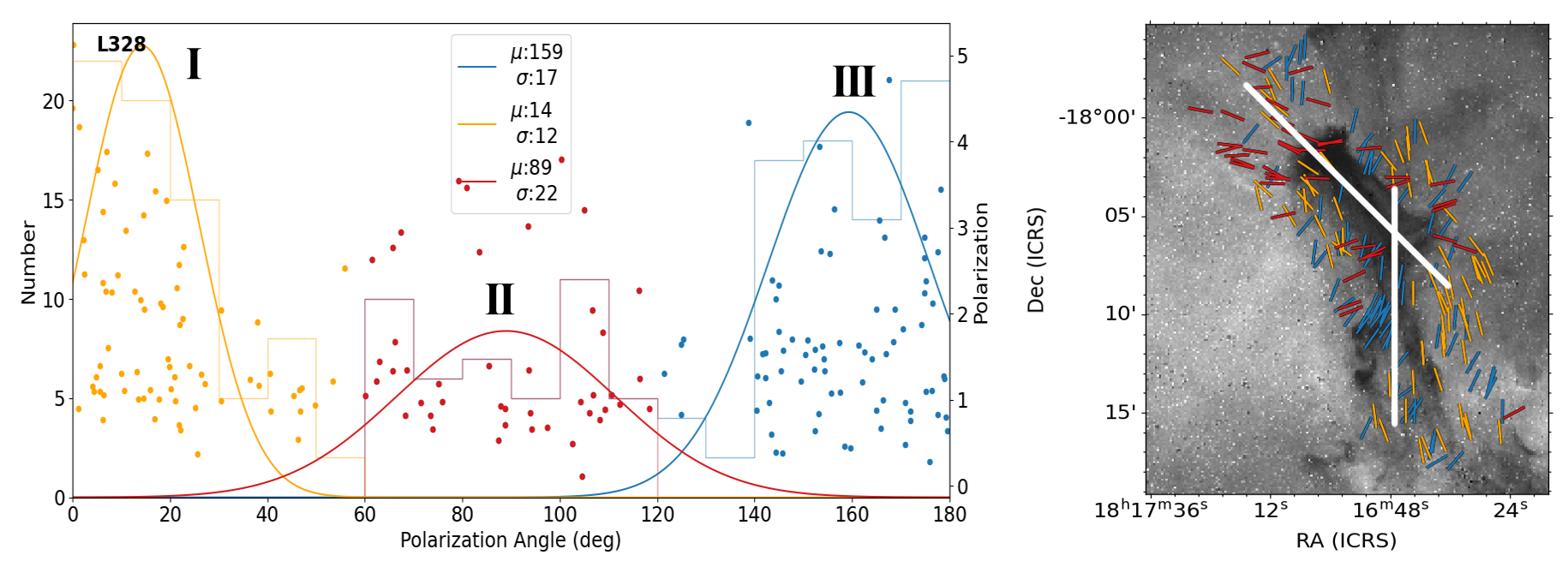}
\caption{The figure above shows the distribution of polarization angles for L328 with a Gaussian fit to the polarization angle distribution on the left. The distribution has been divided into smaller distributions \textbf{I}, \textbf{II} and \textbf{III} colored as yellow, red and blue respectively. The colored dots represent the polarization of stars, where larger dots show higher polarization and vice-versa. On the right we have distributions \textbf{I}, \textbf{II} and \textbf{III} plotted on the cloud. The major axis for the head and tail has been plotted in white.}
\label{fig:hist_L328}
\end{figure*}

\begin{figure*}
\includegraphics[width=1\textwidth]{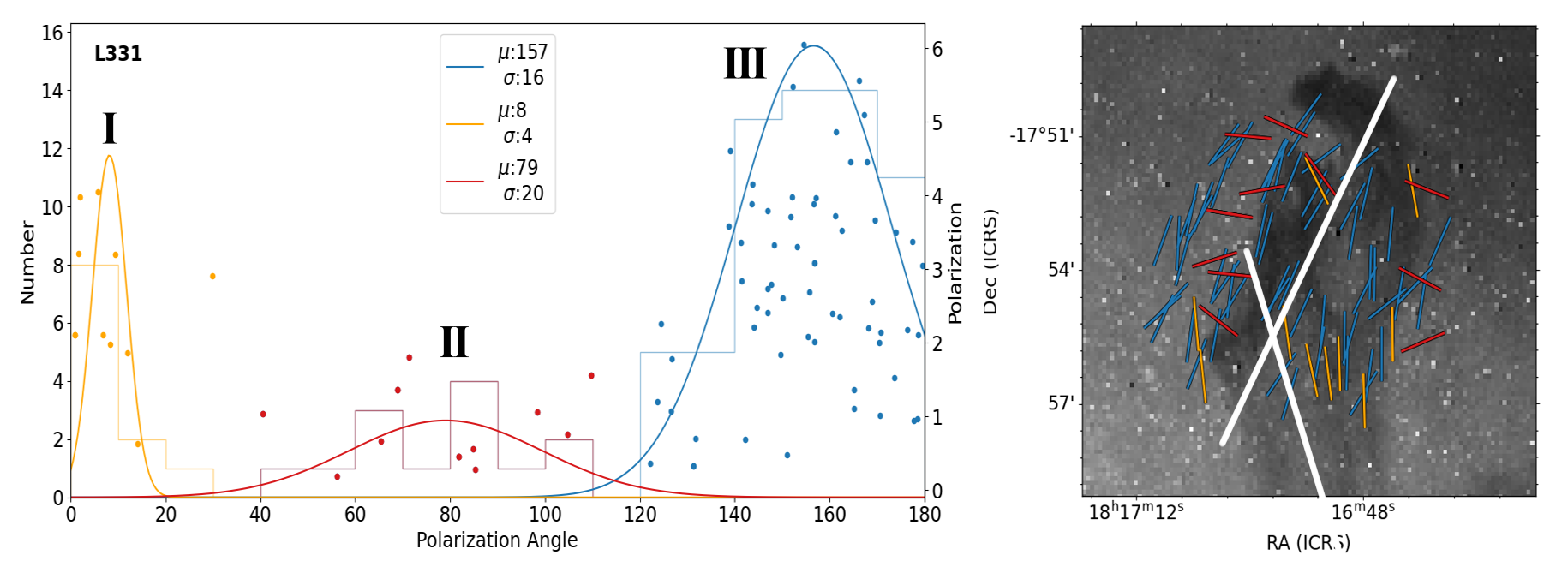}
\caption{The figure above shows the distribution of polarization angles for L331 with a Gaussian fit to the polarization angle distribution on the left. The distribution has been divided into smaller distributions \textbf{I}, \textbf{II} and \textbf{III} colored as yellow, red and blue respectively. The colored dots represent the polarization of stars, where larger dots show higher polarization and vice-versa. On the right we have distributions \textbf{I}, \textbf{II} and \textbf{III} plotted on the cloud. The major axis for the head and tail has been plotted in white.}
\label{fig:hist_L331}
\end{figure*}


Figure \ref{fig:Halpha} shows the R-band subtracted H$\alpha$ image of the cloud morphologies. The clouds have varied morphology even though they are in close proximity to each other. It is a dark, opaque region of $\sim 3' \times 3'$ and several long, curved structures extending to the south-west direction, with size up to $\sim 11'$. The cloud as a whole shows a head-tail morphology, with the tail extending to $\sim 15'$ in projection on the sky. A similar cometary morphology shown by L331, which is located around  10’ to the north of L328, is suggestive of a common external effect on both of these clouds. The cloud's main body has a size of $\sim 3' \times 3'$ and an area of 0.005 sq deg \cite{1962ApJS....7....1L}. The cloud L323, located to the west of L328 at an angular distance of  20’, shows a sharp edge to the east of the globule, again indicating a common possible external influence on the cloud. The cloud has a size of $\sim 11' \times 11'$  and an area of 0.016 sq deg \cite{1962ApJS....7....1L}. All three clouds in the region have the major axis of their heads pointing north-east and the major axis of their tails pointing in the opposite direction, south-west. The cloud morphology is similar to that of cometary globules that are typically found near OB associations in HII regions \cite{1983A&A...117..183R}. The presence of these stars can influence the kinematics and morphology of the clouds using inhomogeneous radiation pressure, forcing the formation of head-tail morphology. This would suggest that UV radiation from some ionizing source in the east of these clouds is affecting their structure. Within this hypothesis, we could explain the absence of prominent tail morphology in L323, as it would lie in the shadow of L328 and thus be shielded from radiation pressure. The three closest B type stars lying within 2$^o$ of the cloud structures were HD 167863, HD 168675, and HD 168352 at 47.5’, 1.11$^o$ and 1.18$^o$ angular separation respectively (see Figure \ref{fig:star_plots}).

\subsection{Magnetic field morphology}
In this work, we use complementary R-band and 850$\mu$m $Planck$ polarization data \citep{2016A&A...594A...1P, 2016A&A...594A..26P} to study the small scale and large scale magnetic field morphology of the region, respectively. These polarization vectors were plotted on an R-band subtracted H$\alpha$ image of the region. Here, the length of the vectors has been normalized for clarity; they do not represent the polarization percentage. The vectors for $Planck$ polarization have been rotated 90$^o$ degrees to map out the magnetic field vectors. The field morphology toward these clouds seen in $Planck$ maps (left panel of figure \ref{fig:plank_optical}) is relatively simple and easy to explain. The overall fields seem to be following a north-south direction away from the clouds. If we concentrate on the L328 cloud and its surroundings, we can see a change in the magnetic field orientations. The magnetic fields follow the cloud morphology, moving from head to tail and being parallel to the major axis of the cloud. This is what you would expect from magnetic fields being frozen inside molecular clouds. The magnetic field again bends and seems to follow the cloud morphology in L323 to some extent, but not as clearly as seen in L328. L331 is found to mostly follow the large-scale field structure and doesn't seem to affect the magnetic fields to the extent of L328 or L323.\\


To understand these field orientations better, we can look at the optical polarization maps (right panel of figure \ref{fig:plank_optical}). L323 has mostly field lines going from north-east to south-west in the tail part, but the lines are totally orthogonal to that when we look at the lower part of the cloud. In the case of L328, the magnetic field lines seem to be following the major axis of the head with a little bit of randomness, and the tail part seems to have been dragging the field lines with the cloud material. The cloud L331 is simpler to explain, but there is a lack of measurements, especially toward the head part. More cannot be said about this cloud except that the field lines mostly follow the cloud morphology. When we compare the large scale magnetic fields from $Planck$ and the small scale magnetic fields from optical polarization maps, they seem to be in agreement with each other.\\

\subsection{Polarization and polarization angle as a function of distance}

The distances of the stars targeted in this study for polarization observations are important as the polarization of the starlight from the targets lying behind the cloud and shining through the cloud material will give information on the contribution of the cloud dust in polarizing the light. Figure \ref{fig:p_vs_d} plots the polarization in optical as a function of distance to the source, and figure \ref{fig:pa_vs_d} shows the polarization angle as a function of distance. The y-axis shows the polarization in percentage (figure \ref{fig:p_vs_d}) or polarization angle in degrees (figure \ref{fig:pa_vs_d}) , and the x-axis shows the distance in kpc adopted from the Gaia EDR3 \cite{2021yCat.1352....0B}. The plots are divided into different colored segments and these segments have the same designated color across the two figures. The polarization angles have been measured from the north increasing towards the east. Only the data points with a ratio of polarization and the error in polarization $\sigma_{P}$, $\frac{P}{\sigma_P} > 2$ were considered for this study. The literature distance to these three clouds is approximately $\sim \hspace{1pt}220\hspace{4pt} pc$ \citep{2011A&A...536A..99M}, the closest data point in our data-set is located at $\sim\hspace{1pt}500\hspace{4pt}pc$ these information implies that all our data points are from stars behind the clouds. In figure \ref{fig:p_vs_d} we observe a common feature present in all three plots. A low polarization sinusoidal pattern oscillating at about 0.5\% polarization and with a half wavelength of about $\sim 0.5-1$ kpc seems to be present in the data for all three cases. This sinusoidal feature is interesting for further investigations, possibly related to the galactic magnetic field. In addition to the sinusoidal feature, L323 has some interesting patterns (see \ref{fig:p_vs_d} \textbf{a} ). There is a sudden increase in polarisation at both 1.2 and 1.9 kpc and a sudden decrease between 1.67 and 1.9 kpc. Up until a sharp increase in polarisation percentage at 2kpc (see figure \ref{fig:p_vs_d} \textbf{b} ), L328's polarisation percentage is around 1$\%$. L331 (\ref{fig:p_vs_d} \textbf{c} ) has very few data points, but there is a faint sinusoidal pattern and a change in polarisation percentage around 2.1 kpc, similar to L328. Even though the distribution of polarisation angles appears chaotic, some observations can be made. All three plots (see \ref{fig:pa_vs_d}) show at least two major distributions around 10$^o$ and 150$^o$, and another minor distributions around 90$^o$. This implies that the polarization angle distribution remains consistent for a long distance behind the clouds.

\subsection{Distribution of polarization angles}

In this section, we look at the distribution of polarization angles and their respective polarizations. When looking at the distribution of P in all three clouds, there doesn't seem to be a pattern. There is a random distribution between polarization and polarization angle in all three clouds. On the other hand, we can make some interesting comments on the distribution of polarization angles. For each cloud, we have plotted an histogram, taking into account the $180^{o}$ ambiguity in the direction of the polarization angles measured. For each cloud, the figure was constructed by subtracting $180^{o}$ from polarization angles above $180^o$ to make the final plot between $0^{o} - 180^{o}$ (Left figures of \ref{fig:hist_L323}, \ref{fig:hist_L328} and \ref{fig:hist_L331}). The histogram distribution was also divided around the major and minor peaks. They have been labelled as I, II, and III and coloured differently. We also put the major axis of the clouds on them for reference (Right figures of \ref{fig:hist_L323}, \ref{fig:hist_L328} and \ref{fig:hist_L331}). For the first part of this section we take a look at the histogram of polarization angle distribution. For L323 (see left image of figure \ref{fig:hist_L323}) the largest peak occurs at $39^{o} \pm 18$ (marked as \textbf{I} in yellow) and the small peak occurs at $140^{o} \pm 32$ (marked as \textbf{II} in blue). This suggests that most of our polarization angles are oriented in a north-east to south-west direction. The plot (see left image of figure \ref{fig:hist_L328}) for L328 has three different peaks, with two major peaks at $14^{o} \pm 12$ (marked as \textbf{I} in yellow) and $159^{o} \pm 17$ (marked as \textbf{III} in blue) and a smaller peak at $89^{o} \pm 22$ (marked as \textbf{II} in red). The two major peaks are at the edges of our range of $0^{o} - 180^{o}$, which suggests that the polarization angles in L328 are approximately oriented in the north-south direction. Similarly (see left image of figure \ref{fig:hist_L331}) L331 has three peaks, with a major peak at $157^{o} \pm 16$ (marked as \textbf{III} in blue) and smaller minor peaks at $8^{o} \pm 4$ (marked as \textbf{I} in yellow) and $79^{o} \pm 20$ (marked as \textbf{II} in red) and the polarization angles in L331 are approximately oriented in the north-south direction. These results suggest that there may be a common underlying physical mechanism responsible for the polarization orientation in both L328 and L331. Further investigation is needed to determine the cause of this alignment and whether it is present in other objects in the same region.\\

For the second part of the analysis, we plotted the divided polarisation angle distributions on their clouds in the same colour as the distribution. For L323 (see right figure of \ref{fig:hist_L323}), the distribution \textbf{II}, is mostly located on the cloud's lower east side and is perpendicular to the major axis of the cloud. The \textbf{I} distribution is mostly concentrated on the upper west side and is parallel to the major axis of the cloud. This suggests that there may be a correlation between the orientation of the polarisation angle distribution and the orientation of the cloud's major axis. In L328 (see right figure of \ref{fig:hist_L328}), the tail part is dominated by distributions \textbf{I} and \textbf{III} and follows the tail structure. The upper part of the head is dominated by distribution \textbf{II} and its parallel to the axis of the head. The main body of the cloud has a more even distribution of all three.  There are not a lot of data points for L331 (see the right figure of \ref{fig:hist_L331}), but a pattern can be observed. The majority of them are in distribution \textbf{III} and are distributed over the body and also follow it. Distribution \textbf{I}, which follows the diffuse tail lie in the lower region of the cloud. Distribution \textbf{II} is mostly perpendicular to the major axis of the body.   The polarisation angles in the three clouds are not yet fully understood due to the sparse sampling of data points. Further studies are needed to provide a more comprehensive understanding of the distribution of polarisation angles in these clouds.

\begin{figure*}
\includegraphics[width=1\textwidth]{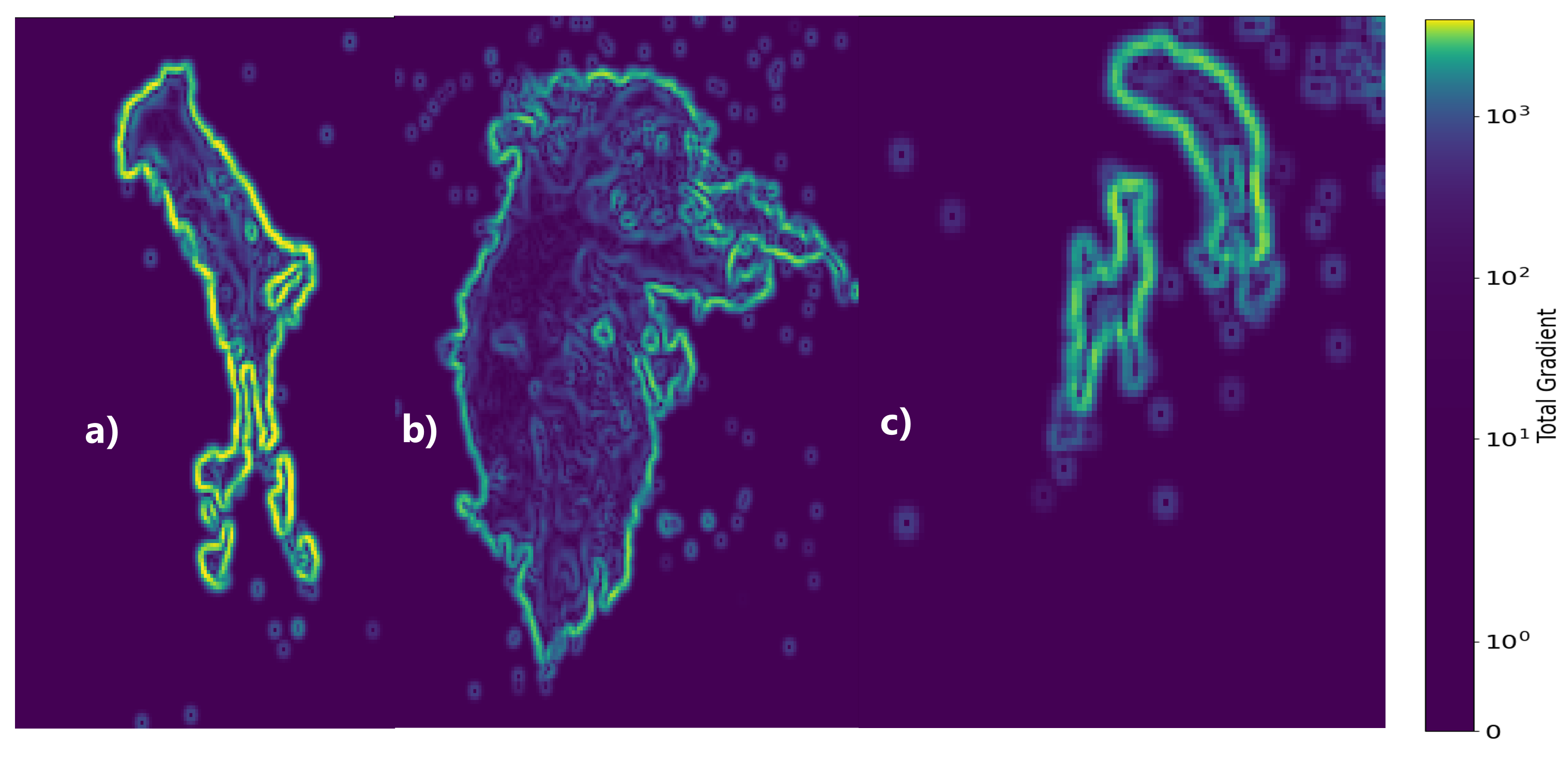}
\caption{The figure above shows the total gradient ($\nabla N_{tot} = (\nabla N_x^2 + \nabla N_y^2)^{1/2}$) which are essentially the edges for the three clouds L328 (a), L323 (b), and L331 (c) (not to scale). The edges were computed using a 1$\sigma$ Gaussian filter truncated at 4$\sigma$ and a Sobel derivative kernel.}
\label{fig:hro_grad}
\end{figure*}

\subsection{Histogram of Relative Orientation analysis}
The histogram of relative orientation is a novel method of quantifying the relationship between magnetic fields and cloud structures introduced by \cite{2013ApJ...774..128S}. It uses information from the gradient in x,y (z) to map the density morphology in 2D(3D) maps to compare the polarization directions ($\hat{E}$) and opacity gradients ($\nabla N$). The method provides a more rigorous characterization of the density field that can be used with polarization observations to investigate the relative orientation of cloud morphology and the magnetic field. A more straightforward way to define this would be to find the relative orientation angle $\phi$ between the polarization vector ($\hat{E}$) and the gradient of the column density structure ($\nabla N$). The HRO is the distribution of the relative orientations $\phi$. In particular, this method describes any preference for parallel or perpendicular alignment of the magnetic field with respect to elongated structures seen in column density maps. On applying this method to synthetic polarization measurements from MHD simulations, it was discovered that the gas structures show preferential alignment with magnetic field orientations depending on physical conditions. In the case of low column densities, the gas structures prefer parallel alignment. With sufficiently high magnetization, this preference changes from parallel to perpendicular at higher column densities. In this paper, we will implement the 2D version of this method with some modifications described in the following sections.

\subsubsection{HRO construction}
For a histogram of relative orientation analysis, a column density map and a projected polarization angle map are required, but for our study, neither of them were available. The R band subtracted H$\alpha$ image was used as an alternative for the density maps, and the optical polarization data was used as an alternative to the polarization map. As edge detection essentially requires an image where the target object stands out clearly from the background, allowing us to quantify the target's morphology, the H$\alpha$ image could be used in place of the column density maps. The R-band optical polarization data was gridded into 6'' resolution grids equal to the H$\alpha$ image resolution to create a projected polarization angle map. The H$\alpha$ image map was then split up into three slices, each containing only a single cloud, so they could be worked separately. The slices were then cleaned up a bit to remove the extra background data points surrounding the clouds using a low pass filter with the cutoff set at 2000 (in the unit of intensity for the fits file) for L328 and 1500 for L331 and L323, respectively. These slices were then convolved with a 1$\sigma$ Gaussian filter truncated at 4 $\sigma$ and then the Sobel derivative kernel was applied to calculate the gradient $\nabla N$ . Figure \ref{fig:hro_grad} shows the gradient map of the clouds created from the above algorithm. After the creation of gradient vectors and the polarization vectors from the polarization map, we computed the relative orientation angle ($\Phi$) using.

\begin{equation}\label{gradient}
\phi = arctan\left( \frac{\nabla N \times \hat{E}}{\nabla N.\hat{E}}\right)
\end{equation}

Under this convention, $\phi = 0^o$ represents parallelism, while $\phi \pm 90^o$ represents perpendicularity between the magnetic field and the cloud morphology. The histogram of these relative orientations is then used to do the HRO analysis for these clouds.  

 \subsubsection{Relative Orientation Parameter}
  
 The most efficient method of representing the characteristic shape of the histogram describing the distribution of $\phi$ is using $\xi$, the normalized version of the HRO shape parameter \cite{2013ApJ...774..128S}. The parameter is defined as  
 
 \begin{equation}\label{error}
\xi = \frac{A_c - A_e}{A_c + A_e}
\end{equation}
 
where the $\xi$ parameter compares the area of the center of the histogram ($A_c$) with the area of the extremes of the histogram ($A_e$). The central area ($A_c$) is the area of the region $-22.5^o < \phi < 22.5^o$ of the histogram, while the extreme area ($A_e$) is defined as the area under the union of the regions $-90^o < \phi < -67.5^o$ and $67.5^o < \phi < 90^o$. Under this definition, $\xi$ can hold values between -1 and 1. Here, $\xi > 0$ indicates a histogram with a peak between $-22.5^o$ and $22.5^o$ which corresponds to a preference for parallel alignment between cloud morphology and the magnetic field. Whereas $\xi < 0$ indicates a histogram where the magnetic field is largely perpendicular to the cloud morphology. In the case of the shape parameter being approximately zero ($\xi \approx 0$), this suggests no alignment preference and the corresponding histogram will be largely flat.\\

The uncertainty in $\xi$ is $\sigma_{\xi}$ defined as
\begin{equation}\label{error_in_sigma}
\sigma_{\xi}^2 = \frac{4(A_e^2\sigma_{A_c}^2 + A_c^2\sigma_{A_e}^2)}{(A_c + A_e)^2}
\end{equation}

where $\sigma^2_{A_e}$ and $\sigma^2_{A_c}$ are the variances of $A_e$ and $A_c$. They can be obtained following $\sigma_k^2 = h_k(1 - h_k/h_{tot})$, where $h_k$ and $h_{tot}$ are numbers of data points in the central or extreme bins and of the total data points, respectively. This uncertainty describes the 'jitter' of the histogram, explained in \citet{2016A&A...586A.138P}. The results of our HRO analysis suggest that L328 has values of -0.328 and 0.060, L323 has values of 0.08 and 0.057, and L331 has values of 0.055 and 0.095 for $\xi$ and $\sigma_{\xi}$, respectively. The results have been summarized in the Table \ref{tab:HRO}.

\begin{table}
\centering
\caption{The table below summarizes the result of HRO analysis}
\label{tab:HRO}
\begin{tabular}{llll}
\hline
Cloud   & $\xi$  & $\sigma_{\xi}$ \\
\hline\hline
L328 & -0.328 & 0.060\\
L323 & 0.08  & 0.057\\
L331 & 0.055 & 0.095\\
\hline\hline
\end{tabular}
\end{table}

\section{Conclusion}

In this work, we presented a plane of the sky projected magnetic field in L328, L323, and L331, obtained using R-band polarimetry and 850$\mu$m $Planck$ polarization maps. We investigated the magnetic and cloud morphologies, distribution of polarization and polarization angles, the magnetic field strength in the clouds using the modified CF method, and the relative orientation of the cloud with the magnetic fields. The main findings of the study are as follows:\\

\begin{enumerate}
    \item  When observed in the R-band, the overall magnetic field structure in L323 is more ordered and parallel to the cloud in the south-western and north-western regions of the cloud and mostly perpendicular in the north-eastern region of the cloud. There is more randomness in the orientation of the magnetic field of L328 and L331. The large-scale magnetic fields inferred from 850$\mu$m $Planck$ polarisation maps are in agreement with the small scale magnetic fields. The agreement between large and small scale magnetic fields indicates that the overall magnetic field structure is consistent across different scales.\\

    \item The Gaussian fit to the distribution of the polarization angle distribution for L323 had values for $\mu$ and $\sigma$ 38$^o$ and 10$^o$. For L328, there were two major peaks with $\mu$ 14$^o$ and 159$^o$ and $\sigma$ 12$^o$ and 16$^o$. Similarly, the fit for L331 had two major peaks with $\mu$ 10$^o$ and 157$^o$ and $\sigma$ 5$^o$ and 16$^o$. \\

    \item The spatial distribution of polarisation angles suggests that three potential ionising sources in the east are influencing a fairly complex magnetic field morphology. This necessitates a more in-depth investigation, possibly at sub-mm/mm wavelength, to capture the emission dust polarisation at a much higher resolution than the $Planck$ survey in order to fully comprehend the overall magnetic field morphology of the area.\\
      
    
    \item Our modified HRO analysis revealed that the magnetic fields are perpendicular to the cloud in L328 and parallel to the cloud in L331 and L323. The orientation in L323 and L331 is close to zero, which suggests that most of the polarization angles have no preference for either parallel or perpendicular orientation.\\

\end{enumerate}

\section*{Acknowledgements}

The authors thanks the referee and the scientific editor for an encouraging report resulting in significant improvement to the paper. We acknowledge the use of Planck and Gaia survey data from European Space Agency mission. We thank the staff's help and equipment contribution in order to carry out the project's observation using the active optical polarimeter, AIMPOL, at the Sampurnanand telescope in ARIES, India.

\section*{DATA AVAILABILITY}
The data used for this study are available from the Gaia and Planck survey from European Space Agency. The polarisation measurement data used for the analysis are available in a machine-readable table as an appendix to this paper, and data products will be shared on reasonable request.




\bibliographystyle{mnras}
\bibliography{mnras_template} 

\appendix
\section{Online Data}
The outcomes of our observations using the active optical polarimeter, AIMPOL, at the Sampurnanand telescope in ARIES, India, are attached in this appendix. Star's Gaia ID, Right ascension ($\alpha$ $^o$), declination ($\delta$ $^o$), polarisation percentage ($P$ $\%$), polarisation angle ($\theta$ $^o$), and Distance (kpc) are the parameters that are displayed in the tables below in six columns. Table \ref{tab:L331_data} contains the sample data for L331, while tables \ref{tab:L323_data1} contains the sample data for L323, and \ref{tab:L328_data1} contains the sample data for L328. The full tables are available online as supplementary materials. The tables provide a comprehensive dataset that can be used for various astronomical studies and research.

\begin{table}
\centering
\caption{The table below shows the sample polarization results of 10 out of 53 stars (with $P/\sigma_{P} \geq 2$) observed in the direction of LDN 331. The full table is available as supplementary material online.}

\label{tab:L331_data}
\begin{tabular}{llllll}
\hline
Gaia ID & $\alpha (J2000)$ & $\delta (J2000)$ & P $\pm$ $\sigma_P $ & $\theta \pm \sigma_{\theta}$ & Distance \\
 & $(^o)$ & $(^o)$ & ($\%$) & $(^o)$ & (kpc)\\\hline
4097176531550832640 & 274.1960 & -17.8545       & 1.9$\pm$0.5          & 12$\pm$8                & 3.180          \\
4097176531550832640 & 274.1960     & -17.8545       & 3.9$\pm$1.4          & 144$\pm$10              & 3.180          \\
4097170552956347904 & 274.1835     & -17.8737       & 5.5$\pm$2.3          & 152$\pm$12              & 2.626          \\
4097176394111840256 & 274.2364     & -17.8487       & 1.0$\pm$0.3          & 40$\pm$8                & 0.414          \\
4097176394111835136 & 274.2381     & -17.8511       & 2.9$\pm$0.7          & 30$\pm$7                & 1.736          \\
4097176256678404608 & 274.2220     & -17.8607       & 2.8$\pm$0.9          & 148$\pm$9               & 3.151          \\
4097176703349465600 & 274.2540     & -17.8452       & 1.4$\pm$0.4          & 165$\pm$8               & 1.781          \\
4097176703349464064 & 274.2558     & -17.8447       & 3.3$\pm$0.7          & 148$\pm$6               & 1.892          \\
4097176222313156096 & 274.2055     & -17.8711       & 1.5$\pm$0.3          & 174$\pm$5               & 0.541          \\
4097176737709193728 & 274.2719     & -17.8376       & 3.8$\pm$0.6          & 147$\pm$5               & 1.789          \\
\end{tabular}
\end{table}

\begin{table}
\centering
\caption{The table below shows the sample polarization results of 10 out of 141 stars (with $P/\sigma_{P} \geq 2$) observed in the direction of LDN 323. The full table is available as supplementary material online.}
\label{tab:L323_data1}
\begin{tabular}{llllll}
\hline
Gaia ID & $\alpha (J2000)$ & $\delta (J2000)$ & P $\pm$ $\sigma_P $ & $\theta \pm \sigma_{\theta}$ & Distance \\
ID & $(^o)$ & $(^o)$ & ($\%$) & $(^o)$ & (kpc)\\\hline
4095651710771999360 & 273.9606 & -18.1943 & 4.7$\pm$0.2 & 25$\pm$1  & 1.707 \\
4095651715085418240 & 273.9696 & -18.1933 & 2.8$\pm$0.4 & 92$\pm$4  & 2.576 \\
4095651062250380672 & 273.9881 & -18.1877 & 1.2$\pm$0.2 & 65$\pm$5  & 1.537 \\
4095650856091941120 & 273.9762 & -18.2138 & 3.8$\pm$0.2 & 179$\pm$2 & 2.158 \\
4095650546854298240 & 273.9606 & -18.2238 & 1.4$\pm$0.2 & 158$\pm$3 & 2.018 \\
4095650508180896896 & 273.9464 & -18.2333 & 3.8$\pm$0.8 & 94$\pm$6  & 1.342 \\
4095650890451673472 & 273.9912 & -18.2075 & 2.4$\pm$0.4 & 129$\pm$4 & 1.719 \\
4095650684293235072 & 273.9948 & -18.2186 & 2.2$\pm$0.4 & 169$\pm$5 & 2.537 \\
4095650340696210944 & 273.9430 & -18.2491 & 2.1$\pm$0.4 & 79$\pm$5  & 2.717 \\
4095650649933496192 & 273.9883 & -18.2254 & 2.0$\pm$0.4 & 158$\pm$5 & 2.839 \\
\end{tabular}
\end{table}

\begin{table}
\centering
\caption{The table below shows the sample polarization results of 10 out of 191 stars (with $P/\sigma_{P} \geq 2$) observed in the direction of LDN 328. The full table is available as supplementary material online.}
\label{tab:L328_data1}
\begin{tabular}{llllll}
\hline
Gaia ID & $\alpha (J2000)$ & $\delta (J2000)$ & P $\pm$ $\sigma_P $ & $\theta \pm \sigma_{\theta}$ & Distance \\
ID & $(^o)$ & $(^o)$ & ($\%$) & $(^o)$ & (kpc)\\\hline
4097156740341337216 & 274.1967 & -17.9962 & 1.2$\pm$0.2 & 179$\pm$5  & 2.815 \\
4097156740341314688 & 274.2022 & -18.0022 & 2.4$\pm$0.1 & 6$\pm$1    & 2.521 \\
4097156671621822080 & 274.2024 & -18.0107 & 1.4$\pm$0.1 & 6$\pm$2    & 1.058 \\
4097155091073885952 & 274.1551 & -18.0389 & 0.4$\pm$0.1 & 144$\pm$8  & 1.318 \\
4097155194153088256 & 274.1675 & -18.0338 & 0.6$\pm$0.3 & 152$\pm$14 & 1.397 \\
4097156671621804928 & 274.2097 & -18.0129 & 1.4$\pm$0.2 & 20$\pm$5   & 1.029 \\
4097155159793327360 & 274.1849 & -18.0300 & 0.8$\pm$0.2 & 17$\pm$7   & 0.954 \\
4097155159793324544 & 274.1858 & -18.0311 & 0.8$\pm$0.2 & 6$\pm$6    & 0.612 \\
4097155125433588864 & 274.1737 & -18.0399 & 3.8$\pm$0.4 & 100$\pm$3  & 2.449 \\
4097156602902288384 & 274.2384 & -18.0062 & 0.9$\pm$0.4 & 140$\pm$12 & 2.133 \\
\end{tabular}
\end{table}

\bsp	
\label{lastpage}
\end{document}